\documentclass[twocolumn,nofootinbib,preprintnumbers,floatfix,superscriptaddress,aps]{revtex4}

\usepackage{amsmath,graphicx,verbatim,epsfig}

\usepackage{epsfig,epstopdf}
\usepackage[utf8x]{inputenc}
\usepackage{amsmath,graphicx}
\usepackage{color}

\newcommand{\la}{\lambda}

\newcommand{\nn}{\nonumber}

\newcommand{\bn}{{\bar n}}

\newcommand{\nslash}{{\not \!n}}

\newcommand{\nb}{\bar n}

\newcommand{\be}{\begin{equation}}
\newcommand{\ee}{\end{equation}}
\newcommand{\bea}{\begin{eqnarray}}
\newcommand{\eea}{\end{eqnarray}}
\newcommand{\balign}{\begin{align}}
\newcommand{\ealign}{\end{align}}

\newcommand{\bg}{\begin{gather}}
\newcommand{\foma}{\end{gather}}

\newcommand{\noopsort}[1]{}

\def\pd{\partial}

\def\L{\Lambda}

\def\<{\langle}
\def\>{\rangle}

\def\d{\delta}  
   \def\L{\Lambda}

\def\m{\mu}

\def\t{\tau}

\def\({\left(}
\def\[{\left[}
\def\){\right)}
\def\]{\right]}

\def\Dslash{D\!\!\!\!\slash}

\def\nslash{n\!\!\!\slash}
\def\bnslash{\bar n\!\!\!\slash}

\def\pdslash{\partial\!\!\!\slash}
\def\Aslash{A\!\!\!\slash}

\begin{document}
\baselineskip 3.0ex

\vspace*{18pt}

\title{SCET, Light-Cone Gauge and the T-Wilson Lines }

\author{Miguel Garc\'ia-Echevarr\'ia}\email{miguel.gechevarria@fis.ucm.es}
\author{Ahmad Idilbi}\email{ahmadidilbi@gmail.com}
\author{Ignazio Scimemi}\email{ignazios@fis.ucm.es}

\affiliation{Departamento de F\'isica Te\'orica II,
Universidad Complutense de Madrid (UCM),
28040 Madrid, Spain}

\begin{abstract}
\baselineskip 3.0ex  \vspace{0.5cm}
Soft-Collinear Effective Theory (SCET) has been formulated since a decade now in covariant gauges.
In this work we derive a modified SCET Lagrangian applicable in both classes of gauges: regular
and singular ones. This extends the range of applicability of SCET. The new Lagrangian must be used
to obtain factorization theorems in cases where the transverse momenta of the particles in the final states are not integrated over, such as
semi-inclusive deep inelastic scattering, Drell-Yan and the Higgs production cross-section at low transverse momentum. By doing so all non-perturbative matrix elements appearing in the factorized cross-sections are gauge invariant.
%This choice of the gauge in fact
 %improves the realization of the collinear symmetry in the theory.  We explain the origin of the new Lagrangian starting from  %QCD.

\end{abstract}
\maketitle
%\section{ Introduction}
%\label{sec:I}
In  re\-cent years Soft-Collinear Ef\-fec\-tive The\-ory (SCET)~\cite{Bauer:2000yr,Bauer:2001ct} has emerged as an important
 tool to describe jet-like events ranging from  heavy quark hadronic decays to Large Hadron Collider (LHC) physics.
The advantage of this effective theory of QCD is that
it incorporates, at the Lagrangian level,  all the kinetic symmetries of a particular jet-like event.
 The fields in SCET are either collinear, anti-collinear or soft (low energetic) depending  whether they carry most 
 of their energy along a
light-like vector ($n,\; n^2=0$), an anti-light-like vector ($\nb,\; \nb^2=0, \;n\nb=2$) or if the energy is soft and radiated
 isotropically.
The current formulation of SCET is rather limited to a class of regular gauges. In this class of gauges the gauge boson
fields vanish at infinity in coordinate space, thus no gauge transformation can be performed at that point.
This limitation has a rather important implications as we discuss below.
Moreover, using a singular gauge like  the light-cone gauge (LCG), it is possible to improve the symmetry of the effective Lagrangian(s) because the gauge fixing conditions also
respect the symmetry of the jet-like event.

In a previous work \cite{Idilbi:2010im} it was argued that by extending the formulation of SCET to the class of singular gauges,
% where the gluon fields do not vanish at the boundary surface and where gauge invariance is not completely obtained,
 a new Wilson line, the $T$-Wilson line, has to be invoked as a basic SCET building block.
The $T$-Wilson line discussed in \cite{Idilbi:2010im}--which is exactly 1 in covariant gauge--is built using LCG ghost field  $A_{n\perp}(x^+,\infty^-,x_\perp)$.
This transverse Wilson line allows for a complete gauge invariant definitions of the non-perturbative matrix elements of collinear particles
to be obtained from first principles,
it allows to \emph{properly} factorize high-energy processes with explicit transverse-momentum dependence and it reads
\footnote{In this paper we  have adapted our convention for Wilson lines to the  one of Ref.~\cite{Bauer:2000yr}. This  is consistent
with  the results of Ref.~\cite{Idilbi:2010im}.},
\begin{align}
\label{eq:T}
 T_{n}%(x^+,x_\perp)=% \nn \\ &
 =\bar P\exp\left[i g\int_0^\infty d\tau  l_\perp \cdot  A_{n\perp}(x^+,\infty^-, x_\perp- l_\perp\tau)
\right]\, ,
 \end{align}
where ${\bar P}$ stands for anti-path ordering.
Below we  discuss also the
transverse Wilson lines built from soft gluons which allows for a gauge invariant definitions of soft matrix elements with transverse space separation.

In this work we show how to implement the $T$-Wilson line at the level of the soft-collinear Lagrangians.
The results obtained (see below) are Lagrangians applicable in covariant gauges as well as in light-cone gauge (LCG) both in SCET-I and  in SCET-II.
This in turn will enable us to explicitly invoke the transverse-momentum dependence to the collinear quark and gluon jets.
Those latter quantities form the fundamental blocks for the non-perturbative matrix elements like transverse-momentum
dependent parton distribution functions (TMDPDFs), beam-functions (BFs) \cite{Stewart:2009yx} and the like~\cite{Mantry:2009qz,Becher:2010tm} .

We start the discussion with the LCG: $\nb A=A^+=0,\; \nb^2=0$.
QCD can be canonically quantized in this gauge~\cite{Bassetto:1984dq} and  the quantization fixes the
Feynman rules for the gauge bosons with the Mandelstam-Leibbrandt (ML) prescription~\cite{Mandelstam:1982cb}.
%The gauge  condition $ A^+=0,\; \nb^2=0$  however does not fix completely the gauge. As a result,
%gauge fields  should include also the so called ``ghost'' bosons, $A(x^+,\infty^-,x_\perp)$~\cite{Bassetto:1984dq}.
Passing from QCD to   SCET  however some  new issues arise.
For instance, in QCD one needs to specify one gauge fixing condition. In the effective theory however every light-cone (LC)  direction defines a collinear gauge sector and it is not clear, beforehand, how the gauge fixing conditions
 %in QCD is ``translated''  into 
respect the power counting of the  different collinear sectors.

In the following we show how LCG can be implemented both in collinear  and soft sectors of SCET. In particular we study  in which cases LCG
is compatible with power counting and ``multipole expansion'' in SCET-I and SCET-II.
First (in sec.~\ref{sec:L}) we illustrate the origin of the $T$-Wilson line from full QCD. Then we explicitly show how the $T$-Wilson line appears
in the SCET Lagrangians.
This result is the main result of this paper: For the (leading order) SCET Lagrangians and the basic building-blocks to be gauge invariant in covariant as well as in LCG, transverse gauge links have to be introduced in SCET-I and in SCET-II.
In sec.~\ref{sec:AC} we discuss the applications of those results then we conclude.

%%%%%%$$$$$$$$$$$$$$$$$$$$$$$$$$$$$$$$$$$$$$$$$$$$$$$$

\section{The QCD and SCET Lagrangians in LC gauge: T-Wilson lines}
\label{sec:L}
In this chapter we want to write the SCET matter Lagrangian in LCG outlining the role of ghost fields in LCG.
First we recall some of the features of the gluon fields in QCD in LCG from Ref.~\cite{Bassetto:1984dq}.
 To fix matters, we work in QCD with the gauge fixing condition $\nb A=0$.
 The canonical quantization of the gluon field proceeds by inserting in the Lagrangian the gauge fixing term
% \be \label{eq:l_gf}
$\mathcal{L}_{gf} = \Lambda^a (\nb A^a) $.
%\ee
 The $\Lambda^a$ is a field whose value on the Hilbert space of physical states is equal to zero.
  It is possible to write the most general solution of the equation of motion of the boson field $A_\mu^a$
 via decomposing it into
  \begin{align} \label{eq:A_bassetto}
A^a_\m(k) &= T^a_\m(k)\d(k^2) + \nb_\m\frac{\d(\nb k)}{k_\perp^2} \L^a(n k,k_\perp)
\nn \\ &
+
\frac{ik_\m}{k_\perp^2} \d(\nb k) U^a(n k,k_\perp)\,,
\end{align}
where the field $T^a_\m$ is such that $\nb^\m T^a_\m(k)=0$ and $k^\m T^a_\m(k)=0$.
 Fourier transforming this expression we see that in general the field $A^a_\m(x)$  has non-vanishing ``$-$''
and ``$\perp$'' (respectively $n A^{a}(x)$ and $ A^{\mu,a}_\perp(x)$) components when $x^-\rightarrow \pm \infty $.
We define
\begin{align}
 A^{(\infty)}  (x^+, x_\perp) &\stackrel{def}{=} A (x^+,\infty^-, x_\perp )\\
{\tilde A}(x^+,x^-, x_\perp) &\stackrel{def}{=} %\nn \\ &&
 A(x^+,x^-, x_\perp) -  A^{(\infty)} (x^+, x_\perp)
 \nn
  %=
%\nn\\
%{\tilde A}_{n &\stackrel{def}{=}  A_{n} -  A_{n}^{(\infty)}
\end{align}
Then we have the following relation:
\be
i\Dslash_\perp = i\pdslash_\perp + g\Aslash_{\perp}=  i\pdslash_\perp +
g\tilde\Aslash_{\perp} + g\Aslash_{\perp}^{(\infty)}  \stackrel{def}{=}
i\tilde\Dslash_\perp  + g\Aslash_{\perp}^{(\infty)}
\ee
 and we  can show that
\be \label{dslash}
i\Dslash_\perp= T i\tilde\Dslash_\perp T^\dagger \,,
\ee
 where
\begin{align}
\label{eq:Tqcd}
 T^\dagger%(x^+,x_\perp)=% \nn \\ &
 =P\exp\left[-i g\int_0^\infty d\tau { l}_\perp \cdot { A}_\perp^{(\infty)}(x^+,{ x}_\perp-{l}_\perp\tau)
\right]\,
 \end{align}
 and  we use that  in Eq.~(\ref{dslash}) the fields $\tilde A_\perp$ and  $A_{\perp}^{(\infty)}$ are evaluated at
 space-like separated points.
The proof of this equation  leads us automatically to the inclusion of the
$T$-Wilson line in the Lagrangian.
%
%We can simplify the equation we want to verify:
%\bea \label{dslash2}
%i\Dslash_\perp \mathcal{O} &=& T_n^\dagger i\tilde\Dslash_\perp T_n \mathcal{O}
%\nn\\
%i\Dslash_\perp \mathcal{O} &=& T_n^\dagger \[i\tilde\Dslash_\perp T_n\] \mathcal{O} +
%T_n^\dagger T_n \[i\tilde\Dslash_\perp \mathcal{O}\]
%\nn\\
%i\pd^\m_\perp + gA^{(\infty)\m}_{n\perp} &=& T_n^\dagger \[i\pd^\m_\perp\] T_n  +
%T_n^\dagger T_n \[i\pd^\m_\perp \]
%\nn\\
%-ig T_n A^{(\infty)\m}_{n\perp} &=& \pd_\perp^\m T_n
%\eea
We write
\begin{align} \label{demo1}
i l_\perp \pd_\perp T^\dagger &= iT^\dagger l_\perp\pd_\perp \[
-ig\int_0^{\infty} d\t  l_\perp  A_{\perp}^{(\infty)}(x^+, x_\perp -
\t  l_\perp) \]
\nn
\\
&=
g  T^\dagger \int_0^\infty d\t \int \frac{d^4k}{(2\pi)^4}
(i l_\perp k_\perp)\,  l_\perp  A_{\perp}^{(\infty)} (k)
\nn \\
&\times
e^{i k (x-l_\perp \tau)}%{-i\left([k^+]\infty^-+k^-x^+ -\veccp{\scriptstyle k} (\veccp{\scriptstyle x}-
%\t \veccp{\scriptstyle l})\right)}\,
%\nn
%\\&
=g T^\dagger l_\perp A_{\perp}^{(\infty)} (x^+, x_\perp) ,
%-ig  T^\dagger \int \frac{d^4k}{(2\pi)^4} e^{-ikx} \frac{k^\m_\perp}{ l_\perp  k_\perp}
% l_\perp  A_{\perp}^{(\infty)}(k)
\end{align}
which proves Eq.~\eqref{dslash}, as $l_\perp$ is an arbitrary vector.
Moreover
$1/\bn\pd$ and $T$ commute because the $T$-Wilson line does not depend on $x^-$. Under gauge transformation
$\delta A_{\mu\perp}^{(\infty)}=D_{\mu\perp}\omega$   one has
$T(x^+, x_\perp)\rightarrow U(x^+, x_\perp) T(x^+, x_\perp) U^\dagger(x^+, x_\perp-l_\perp\infty)= U(x^+, x_\perp) T(x^+, x_\perp)$
 since $A_{\mu\perp}^{(\infty)}(x^+, \infty_\perp)=0 $. Notice also that the $T$-Wilson lines are  independent of $l_\perp$.

Now we split the fermion field into big and small components
 using the  usual projectors $ \nslash \bnslash/4$  and $ \bnslash \nslash/4$ and eliminate the small components using the equations of
motion~\cite{Bauer:2000yr}.
The result of this  is
\begin{align}
\label{eq:LLCG1}
\mathcal{L} &= \bar \xi_n \left( inD + i\Dslash_\perp \frac{1}{i\bn D}
i\Dslash_\perp \right) \frac{\bnslash}{2}\xi_n\ . %\nn \\ &
\end{align}
In QCD in LCG with the gauge condition $\bn A=0$,
\begin{align}
\label{eq:LLCG2}
\mathcal{L} &=
\bar \xi_n \left( inD + T i\tilde\Dslash_\perp
\frac{1}{i\bn \pd} i\tilde\Dslash_\perp T^\dagger \right)\frac{\bnslash}{2} \xi_n\,.
\end{align}
 In order to get the SCET Lagrangian we must implement multipole expansion and power counting on the fields that appear
in Eq.~(\ref{eq:LLCG1}). In SCET we have also the freedom to choose a different gauge in the different sectors of the theory.
 We distinguish the cases of SCET-I and SCET-II.  The two formulations differ essentially in the scaling of the soft sector of the
 theory.
In SCET-I,  collinear fields describe particles whose momentum $k$  scales like $(\nb k, n k, k_\perp)\sim Q(1,\la^2,\la)$ where
$\la\ll 1$ and $Q$ is a hard energy scale. Also the  components of the
collinear gluons, $A_n^\mu$, in SCET-I   $(\nb A_n,n A_n, A_{n\perp})$ scale like $\sim (1,\la^2,\la)$.
The scaling  of ultra-soft (u-soft) momenta in  SCET-I is $\sim Q(\la^2,\la^2,\la^2)$ and u-soft gluon fields 
$(\nb A_{us},n A_{us}, A_{us\perp})$ scale as $\sim (\la^2,\la^2,\la^2)$.
 In SCET-II we have for collinear field: $(\nb k, n k, k_\perp)\sim Q(1,\eta^2 ,\eta)$ where $\eta\ll 1$.
 For soft momentum:
$(\nb k, n k, k_\perp)\sim Q(\eta,\eta,\eta)$  and collinear (soft) gluons field components scale accordingly.
The main difference is that in SCET-I all the components of the
 soft  momentum scale like the
small component of the  collinear fields, while in  SCET-II   the soft
modes scale like the transverse component of the collinear  fields.
%%%%%%%%%%%%%%%%%%%%%%%%%%%%%%%%%%%%%%%%%%%%%%%%%%%%%%%%%%%%%%%%%%%%%%%%%%%%%%%%%%%%%%%%%%%%%%%%%%%%%%%%%%%

\underline{\textit {\bf SCET-I.}}
 Consider first the case  of  SCET-I  where  the u-soft sector is treated in covariant gauge while the LCG is  imposed
  on the collinear  gauge fields.
In LCG: $\bn A_n=0$ and with the multipole expansion~\cite{Bauer:2000yr} of Eq.~(\ref{eq:LLCG1}) gives
\begin{align}
\label{eq:SCETI}
\mathcal{L_I} &=
\bar \xi_n \Big( inD_n + g n A_{us}(x^+)+ T_n i\tilde\Dslash_{n\perp}
\frac{1}{i\bn \pd} i\tilde\Dslash_{n\perp} T_n^\dagger \Big) \frac{\bnslash}{2}\xi_n\,,
\end{align}
 where $iD^{\mu}_n=i\pd^\mu+g A_n^\mu$, the   u-soft field depends on $x^+$
(transverse  and  collinear variations of the u-soft field are power suppressed) and  the $T$-Wilson line is given in Eq.~\eqref{eq:T}.
%\begin{align}
%\label{eq:Tcoll}
% T_n %(x^+,x_\perp)=% \nn \\ &
% =\bar P\exp\left[i g\int_0^\infty d\tau { l}_\perp \cdot { A_n}_\perp^{(\infty)}(x^+,{ x}_\perp-{ l}_\perp\tau)
%\right]\, .
% \end{align}
The presence of the $T$-Wilson line is essential to have gauge invariance as  was shown in Ref.~\cite{Idilbi:2010im}.
Let us  decide  now to impose LCG  also on the u-soft sector  of the  theory, $n A_{us}=0$.  The corresponding
$T$-Wilson line that would  arise, following Eq.~(\ref{eq:LLCG1}) disappears due to multipole expansion and power counting.
 In fact u-soft fields cannot depend on transverse  coordinates in the leading order Lagrangian of SCET-I.
 In other words the $T$-Wilson line for u-soft fields breaks the power counting of SCET-I and so  the u-soft part of
SCET-I cannot be written in LCG. The other possible choice $\bn A_{us}=0$ has no impact on the leading order SCET Lagrangian.
%The situation is even clearer if we use  the same LCG fixing in the collinear sector and in the u-soft sector. Then one would have 
% that Eq.~(\ref{eq:Tqcd}) is
%\begin{align}
%\label{eq:Tcolsoft}
% T %(x^+,x_\perp)=% \nn \\ &
% & ={\bar P}\exp\left[i g\int_0^\infty d\tau\right. \nn \\
%\times&\left. { l}_\perp \cdot ({ A_{n\perp}}+{ A_{s\perp}})(x^+,\infty^-;{ x}_\perp-{ l}_\perp\tau)
%\right]
%\end{align}
% and $T\simeq T_n$ because $A_s$ is suppressed   in the power counting  with respect to $A_n$.
%The absence of the u-soft fields in the final formulation of the $T$-Wilson line makes impossible the  realization of LCG in the u-soft sector.
Thus  the most general formula for the SCET-I Lagrangian is ($W^T_n=T_n W_n$)
\begin{align}
\label{eq:SCETIi}
\mathcal{L_I} &=
\bar \xi_n \Big( inD_n + g n A_{us}(x^+)
\nn \\
&
+
 i\Dslash_{n\perp}W_n^T
\frac{1}{i\bn \pd}  W_n^{T\dagger} i\Dslash_{n\perp}  \Big)\frac{\bnslash}{2} \xi_n\, .
\end{align}
%%%%%%%%%%%%%%%%%%%%%%%%%%%%%%%%%%%%%%%%%%%%%%%%%%%%%%%%%%%%%%%%%%%%%%%%%%%%%%%%%%%%%%%%%%%%%

\underline{\textit {\bf SCET-II.}}
The analysis of the collinear  sector in SCET-II is the same as for SCET-I.
In the soft sector however one  has new features. In regular gauges soft particles  do not interact with collinear  particles because
the interactions knock the collinear fields off-shell. This is also true in LCG except when
 one makes the choice $n A_s=0$ (take  here a covariant gauge for collinear fields for fixing ideas).
 It is  easy  to be convinced that interactions like $\prod_i \phi_n^i(x)A_{s\perp}^\infty (x^-,x^\perp)$, where here ``$\infty$'' refers to the $+$ direction and
$\phi_n^i(x)$ are generic collinear fields, preserve the on-shellness of the collinear particles.
Using multipole expansion the  vertex  becomes $\prod_i \phi_n^i(x)A_{s\perp}^\infty (0^-,x^\perp)$
(because for collinear fields $x^-\sim 1$ and for the soft field $x^-\sim 1/\eta$.)
In this gauge the covariant derivative for collinear particles  becomes
 $i D^\mu=i\partial^\mu+g A^\mu_n(x)+g A_{s\perp}^{(\infty)\mu}(0^-,x_\perp)$. The gauge ghost $A_{s\perp}^{(\infty)}$ however can be decoupled
 from collinear gluons defining a ``soft free'' collinear gluon $A^{(0)\mu}_n(x)= T_{s n} (x_\perp)A^\mu_n(x)T_{s n}^\dagger(x_\perp)$ where
\begin{align}
\label{eq:Tsoft}
 T_{s n} %(x^+,x_\perp)=% \nn \\ &
 =\bar P\exp\left[i g\int_0^\infty d\tau { l}_\perp \cdot { A_s}_\perp^{(\infty)}(0^-,{ x}_\perp-{ l}_\perp\tau)
\right]\, .
\end{align}
Defining $D^{(0)\mu}_n=i\partial^\mu+g A^{(0)\mu}_n $  we  have  $il_\perp D_{\perp}=T_{s n} (x_\perp)il_\perp D_{n\perp}^{(0)}T_{s n}^\dagger (x_\perp) $ and
\begin{align}
\label{eq:SCETIii}
\mathcal{L_{II}} &=
\bar \xi_{n}^{(0)} \Big( in D_{ n}^{(0)}
%\right. \nn \\
%&\left.
 +i\Dslash_{ n\perp}^{(0)}W_{n}^{T(0)}
\frac{1}{i\bn \partial}  W_{n}^{T(0)\dagger} i\Dslash_{ n\perp}^{(0)}  \Big)\frac{\bnslash}{2} \xi_{ n}^{(0)},
\end{align}
where $\xi_{ n}^{(0)}= T_{s n} (x_\perp) \xi_n(x)$ and $ W_{n}^{T(0)}=T_n^{(0)} W^{(0)}$ are made out of soft free gluons.
Thus thanks to $T_{s n}$ Wilson lines the soft
particles are completely decoupled from collinear particles.
The details of the   contemporary gauge choice $\bn A_n=0$  and $n A_s=0$  is left for future
publication~\cite{nos}.

 Alternatevely the Wilson lines in SCET  are obtained  through  ``auxiliary fields'' as  illustrated, in covariant gauge,
in the second work of \cite{Bauer:2000yr}.
 Through a two-step of matching of the full QCD heavy-to-light current onto the SCET-II effective one,
it was shown that the collinear Wilson line and the soft one can be obtained after integrating out all off-shellnesses
larger than $Q^2\eta^2$ by introducing auxiliary fields for each field that goes off-shell due to soft-collinear interactions.
In LCG however the situation is a bit different.
In the case of $T_{n({\bar n})}$ the auxiliary field method would contain a sub-leading
(${\cal O}(\lambda$)) Lagrangian (including only the transverse component of the collinear gluon field.)
Moreover the resulting fermion (in LCG ${\bar n}A_n (nA_{\bn})=0$) is still on-shell due to $\delta({\bar n}k)(\delta(nk))=0$
so the auxiliary field method breaks down. In SCET-II however all soft gluon field components have the same scaling and
in LCG $(nA_{s}({\bar n}A_{s})=0)$ the resulting fermion is off-shell of ${\cal O}(Q^2\eta)\gg {\cal O}(Q^2\eta^2)$
thus the auxiliary field method might be  applicable. More on this in Ref.~\cite{nos}.

%%%%%%%%%%%%%%%%%%%%%%%%%%%%%%%%%%%%%%%%%%%%%%%%%%%%%%%%%%%%%%%%%%%%%%%%%%%%%%%%%%%%%%%%%%%%%%%%%%%%%%%%%%%%%%%%%
\section{\bf Applications and  conclusions}
\label{sec:AC}
The above derived Lagrangians extend the formulation of SCET valid in covariant gauges
to singular gauges as well.
As it is the case in SCET in covariant gauges, the most important application of these
 Lagrangians is establishing factorization theorems for high-energy processes.
This is especially true for differential cross-sections with $p_T$ dependence where
one expects that the non-perturbative matrix elements entering those factorization theorems
to be un-integrated with respect to the transverse momentum.
In such cases those matrix elements need a gauge link in the transverse space so as
to obtain complete gauge invariance. This is implemented \emph{naturally}
with the $T$-Wilson lines that should be invoked at the level of the basic building blocks
of SCET.  The above discussion  allows us to obtain gauge invariant expressions for any non-perturbative matrix elements
involving quantum fields separated in the transverse direction.
The gauge invariant quark jet was given first in \cite{Idilbi:2010im} and is obtained by simply replacing $W_n$
with $ W_n^T=T_n W_n$,
$\chi_n(x)= W_n^{T\dagger} \xi_n(x) $.
Similarly the gluon jet~\cite{Marcantonini:2008qn} reads
\begin{align}
\label{eq:gj}
g {\cal B}_{n\perp}^\mu=[ W_n^{T\dagger} i D_{n\perp}^\mu W_n^T] ,
\end{align}
where the
derivative operator acts only within the square brackets.
These jets enter the different beam functions introduced in Ref.~\cite{Becher:2010tm} and
in \cite{Mantry:2009qz}. In both of these works a low transverse-momentum dependent cross-sections is
considered  respectively for the Drell-Yan production, Ref.~\cite{Becher:2010tm}, and for the Higgs boson production, Ref.~\cite{Mantry:2009qz},
and
the factorization theorems are obtained within the SCET formalism.  However the non-perturbative matrix elements,
the so-called ``beam functions'', entering those factorization theorems (see Eq.~(9) in \cite{Becher:2010tm}
and Eqs.~(32) and (33) in~\cite{Mantry:2009qz})  are gauge invariant only in the class of regular gauges.
As in the case of the TMDPDF in Ref.~\cite{Idilbi:2010im}, the introduction of the $T$-Wilson line at the level of the SCET Lagrangian and the
quark and gluon jets allows us to obtain, from first principles of the SCET, a well-defined and gauge invariant physical
quantities that are relevant for such important LHC processes and cross-sections.
The  gluon jet functions in Ref.~\cite{Mantry:2009qz} 
%reads
%\begin{align}
%&J_n^{\alpha\beta}(\omega,x^+,x_\perp)=  \nn \\
%&\sum_{pols} \langle P \vert [g{\cal B}^{\alpha}_{n\perp}(x^+,x_\perp)
%\delta(\overline{{\cal P}}_n-\omega)g{\cal B}^{\beta}_{n\perp}(0)]\vert P \rangle
%\end{align}
%where 
 should be written  in terms  of ${\cal B}_{n\perp}^\mu$
 as  given in Eq.~\eqref{eq:gj}.

We remark that the dependence of the soft functions on transverse fields and  transverse coordinates is sensible only in the framework of
SCET-II.
The proper definition of the TMDPDF has to include a subtracted soft function to avoid double counting.
 This subtraction has long been argued by Collins and Hautmann \cite{Collins:2000gd,Collins:1999dz} and more recently
verified in \cite{Cherednikov:2007tw} by considering the renormalization group (RG) properties of the TMDPDF.
This soft function includes a transverse gauge link so as to obtain gauge invariance. The analysis
of~\cite{Collins:2000gd,Collins:1999dz,Cherednikov:2007tw} was performed in QCD.
For SCET to generate the results of QCD for the TMDPDF (and for the generalized quantities thereof such as beam functions)
the effective theory formalism has to include quantities like transverse ``soft'' Wilson lines to properly account the gauge invariance
of the soft function and the RG properties of TMDPDFs and beam functions alike. In SCET-II the typical regular gauge matrix element of soft jets
 $< S_n(x)  S_{\bar n}^\dagger(x) S_{\bar n}(0) S_{ n}^\dagger(0) >$  should be replaced by
$< S_n^T(x)  S_{\bar n}^{T\dagger}(x) S_{\bar n}^T(0) S_{ n}^{T\dagger}(0) >$  where $S_{n(\nb)}^T(x)=S_{n(\nb)}(x)T_{s n(s\nb)}(x_\perp)$.
$T_{s n(s\nb)}(x_\perp)$ are the soft Wilson lines that arise with the gauge fixing $n A_s (\bn A_s)=0$ and are 1 otherwise. For instance    fixing $n A_s=0$ one gets
$< T_{sn}(x_\perp)  S_{\bar n}^\dagger(x) S_{\bar n}(0) T_{ sn}^\dagger(0) >$.

Concluding, we have explained the origin of transverse gauge links $T$ within  SCET. We have studied which light-cone conditions are compatible
with the power counting of the effective theory. The relevance
of the $T$-Wilson lines in SCET was studied in Ref.~\cite{Idilbi:2010im}  for TMDPDF. We have   provided a  gauge invariant definition of
 both quark and gluon jets. The $T$-Wilson lines enter the definitions of  non-perturbative matrix elements in some
processes relevant for LHC, like Drell-Yan and Higgs production. We have finally commented on the transverse momentum dependence
 of the soft functions. Further study in  this direction is expected in the near future~\cite{nos}.

%
%%%%%%%%%%%%%%%%%%%%%%%%%%%%%%%%%%%%%%%%%%%%%%%%%%%%%%%%%%%%%%%%%%%%%%%%%%%%%%%%%%%%%%%%%%%%%%%%%%%%%%%%%%%%%%%%%
%\section*{Aknowledgements}
We thank Iain Stewart for his helpful remarks.
This work is supported by  Spanish MEC, FPA2008-00592.  M.G.E. is supported by the PhD funding program of the Basque  Country Government.
A.I. is supported by the Spanish grant CPAN-ingenio
2010. I.S. is supported by Ram\'on y Cajal Program.

\end{document}